\documentclass[aps,prl,twocolumn,showpacs,superscriptaddress,footinbib]{revtex4-1}
\usepackage{amssymb,amsbsy,graphicx,times,subfigure,marvosym,color,bm,hyperref}
\usepackage[latin1]{inputenc}
\usepackage{epstopdf}

\usepackage{hyperref}
\hypersetup{
    colorlinks,%
    citecolor=blue,%
    filecolor=black,%
    linkcolor=blue,%
    urlcolor=blue
}
\begin{document}

\title{Supra-oscillatory critical temperature dependence of Nb-Ho bilayers}
\author{F. Chiodi}
\email{francesca.chiodi@u-psud.fr}
\altaffiliation{also at Institut d'Electronique Fondamentale, CNRS-Université Paris-Sud, 91405 Orsay, France}
\affiliation{{\mbox{Department of Materials Science and Metallurgy,
University of Cambridge, Pembroke Street, Cambridge CB2 3QZ, United
Kingdom}}}
\author{J. D. S. Witt}
\affiliation{{School of Physics and Astronomy, E. C. Stoner Laboratory, University of Leeds, Leeds LS2 9JT, United Kingdom}}
\author{R. G. J. Smits}
\author{L. Qu}
\author {G. B. Hal\'{a}sz}
\affiliation{{\mbox{Department of Materials Science and Metallurgy,
University of Cambridge, Pembroke Street, Cambridge CB2 3QZ, United
Kingdom}}}

\author{\mbox{C.-T. Wu}} 
\author {O. T. Valls} 
\altaffiliation{Also at Minnesota Supercomputer Institute, University of Minnesota,
Minneapolis, Minnesota 55455}
\affiliation{{School of Physics and Astronomy, University of Minnesota, Minneapolis, Minnesota 55455}}

\author{K. Halterman}
\affiliation{{Michelson Lab, Physics Division, Naval Air Warfare Center, China Lake, California 93555}}

\author{J. W. A. Robinson} 
\email{jjr33@cam.ac.uk}
\author{M. G. Blamire} 
\affiliation{{\mbox{Department of Materials Science and Metallurgy,
University of Cambridge, Pembroke Street, Cambridge CB2 3QZ, United
Kingdom}}}

%\date{\today}
\begin{abstract}
We investigate the critical temperature $T_c$ of a thin s-wave superconductor (Nb) proximity coupled to a helical rare earth ferromagnet (Ho). As a function of the Ho layer thickness, we observe {\itshape{multiple}} oscillations of $T_c$ superimposed on a slow decay, that we attribute to the influence of the Ho on the Nb proximity effect. Because of Ho inhomogeneous magnetization, singlet and triplet pair correlations are present in the bilayers. We take both into consideration when solving the self consistent Bogoliubov-de Gennes equations, and we observe a reasonable agreement. We also observe non-trivial transitions into the superconducting state, the zero resistance state being attained after two successive transitions which appear to be associated with the magnetic structure of Ho. 

\end{abstract}
\maketitle

\emph{Introduction --} In recent years, the interest in superconductor-ferromagnet (S-F) hybrids has greatly intensified, partly due to a number of groups reporting the existence of long-range spin polarized supercurrents \cite{Keizer,Birge,JasonSci,aarts,Sosnin}. It has been known for many years that superconducting correlations can penetrate from a superconductor into a non-superconducting non-magnetic metal (N) via Andreev reflection at the S-N interface, where Cooper pairs penetrate the normal metal N as coherent electron-hole Andreev pairs. As long as the phase coherence between the electron and the hole is preserved, the normal metal will exhibit superconducting features, such as the ability to carry a dissipationless supercurrent. At low temperatures ($T \sim 1\,$K), such a supercurrent can penetrate as far as $1\,\mu$m into a clean normal metal. In a S-F junction, the penetration of superconducting correlations into the F is much shorter ranged (typically less than $10\,$nm even for weak F's).  
Two mechanisms account for this: the dephasing effect of the vector potential due to the internal magnetic field (orbital effect); and the exchange interaction, which acts to impose a parallel alignment of the spins in the Andreev pairs. This parallel alignment is at odds with the Andreev pairs' singlet state, which requires antiparallel spins. 
Due to the energy splitting between the spin-up and spin-down conduction bands in the F, the singlet pair amplitude oscillates in position within the F. This induces oscillations in the supercurrent of a S-F-S Josephson junction and in the critical temperature ($T_c$) of a S-F bilayer. In experiments, many oscillation periods have been observed in the supercurrent of S-F-S junctions \cite{Marco,Ryazanov,JasonSFS}; however, S-F bilayers only exhibited reentrant superconductivity followed by one oscillation and then saturation \cite{Jiang,Muhge,Obi,Zdravko,Wong}. In this Letter we report multiple $T_c$ oscillations with $d_F$ in Nb-Ho systems, which cannot be explained within the standard singlet pairing framework and that we attribute to the influence of the inhomogeneous Ho magnetisation on the Nb proximity effect.

It was recently found that spin-polarized triplet Cooper pairs, in which the spins align parallel rather than antiparallel, can be generated by coupling an s-wave superconductor to an inhomogeneous F \cite{Bergeret,Kadigrobov,Houzet}. This striking possibility opens up a whole new field of research, in which superconductivity and ferromagnetism act cooperatively to create long-range, spin polarized, dissipation-free currents. Of most relevance to this paper, the intrinsic conical magnetism of Ho has been shown to generate a triplet supercurrent which was able to penetrate a 16-nm-thick film of Co \cite{JasonSci}. Here we report experiments to examine the influence of the triplet pair correlations on the proximity effect of a thin s-wave superconductor. We measure the critical temperature $T_c$ of Nb-Ho thin film bilayers whilst varying the Ho layer thickness $d_{Ho}$ in the $1.1-15\,$nm range. We observe multiple  $T_c$ vs $d_{Ho}$ oscillations superimposed on a decaying background. These arise from both the spin-splitting effect of the exchange field and from its inhomogeneity, responsible for the spin-polarized triplet correlations. Furthermore, the zero resistance state in the $R(T)$ curves is attained after two separate sharp transitions for certain Ho layer thicknesses. This suggests an extreme sensitivity of the proximity effect to the details of the Ho magnetic structure. 

\begin{figure}[b]
\begin{center}
\includegraphics[width=8cm]{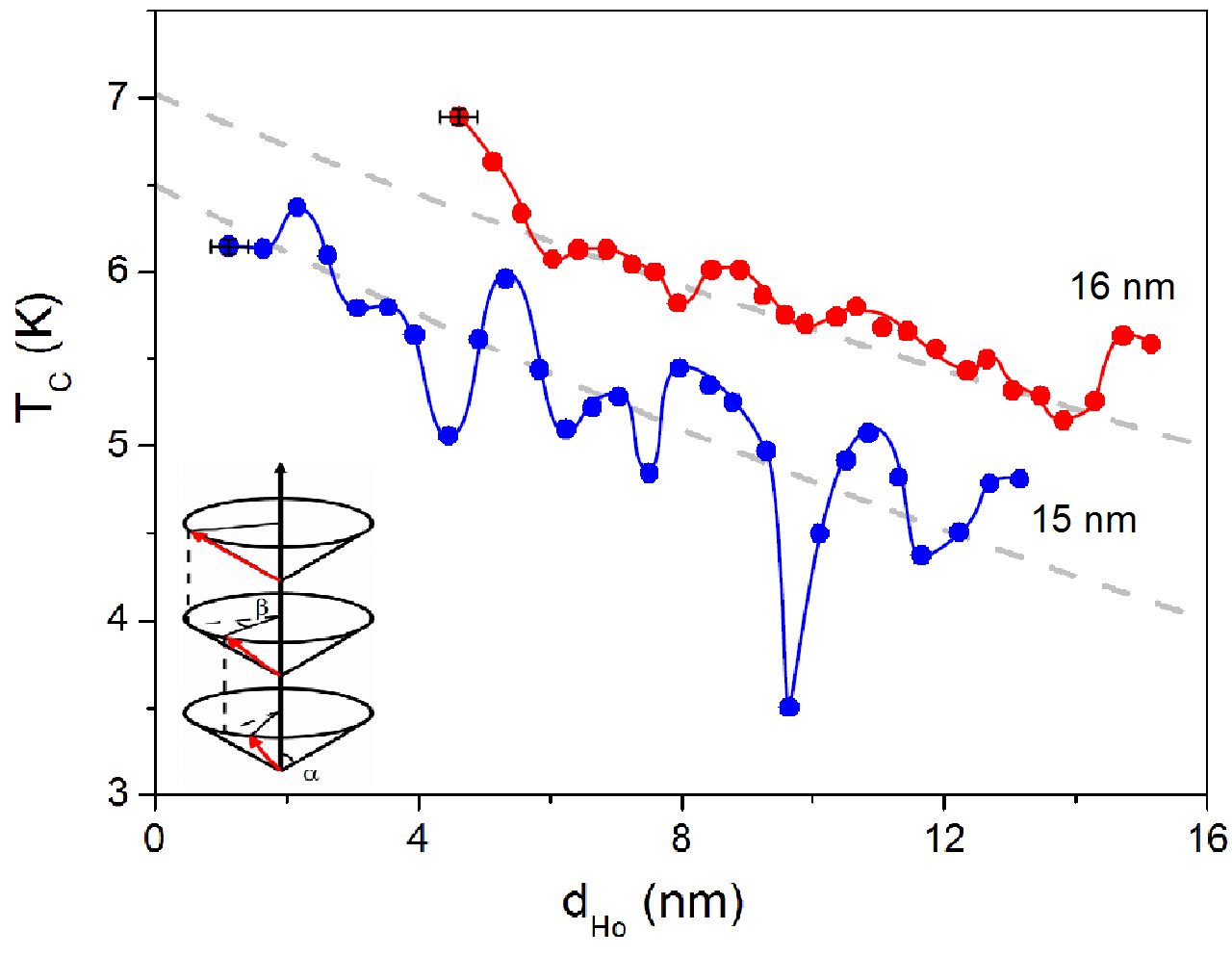}
\caption{(Color online) Superconducting critical temperature $T_{c,50}$ vs Ho thickness $d_{Ho}$ of the T1 series (red, $d_s=16\,$nm) and the T2 series (blue, $d_s=15\,$nm). The oscillations are superimposed on an exponential decay (dotted lines). The error on $T_{c,50}$ is below $\pm 50\,$mK (smaller than the dot), while the error on the Ho layer thickness is $\pm 0.3\,$nm; the error bar, which is the same for all the points, is shown as an example for the first point in each series. The transition for the T2 sample with $d_{Ho}=9.6\,$nm is not complete ($T_{c,50}<4.2\,$K); however, the beginning of the transition being visible, we can estimate $3\,$K$<T_{c,50}<4.2\,$K. Inset: Helical magnetic structure of bulk Ho. }
\label{Tc}
\end{center}
\end{figure}

\emph{Samples and methods --} Holmium is a rare-earth  ferromagnet, which possesses a helical magnetic structure below $133\,$K. At low temperature ($T<19\,$K), Ho is a spiral conical ferromagnet: its atomic moment of 10.34$\,\mu_B$ forms an angle $\alpha=80^{\circ}$ with the crystallographic c-axis which is perpendicular to the sample plane, thus having a small constant out-of-plane component. Its in-plane component rotates along the c-axis, making a helix with an average turn angle of $\beta = 30^{\circ}$ per atomic layer, giving a helical period of $\lambda=3.34\,$nm (see fig.~\ref{Tc}, inset). Thin polycrystalline films of cold-sputtered Ho present a primary (002) phase, the c-axis being normal to the sample surface, and a secondary (101) phase \cite{James}. The Ho helical behaviour is thus relevant for the proximity effect even in polycrystalline films, and is clearly required for the generation of triplet supercurrents in SFS junctions \cite{JasonSci}. 
%but the presence of the secondary phase may influence the magnetic repeat distance and thus the $T_c$ oscillations periodicity, which is determined by both the ferromagnetic coherence length and the Ho helical period. 
The magnetisation loops show a behaviour qualitatively similar to that of a bulk Ho sample, but with a lower saturation moment, suggesting that significant magnetic inhomogeneity is present in sputtered polycrystalline films \cite{supp}. %The growth and magnetic properties of similar polycrystalline films have been reported elsewhere \cite{James}.\\
Two series of Nb-Ho-Cu trilayer thin films were prepared (labeled here as T1 and T2). The Nb and Cu thicknesses were kept constant, while the Ho layer thickness was varied between samples. The T1 and T2 structures are: Nb(16)-Ho[4.4-15]-Cu(20) and Nb(15)-Ho[1.1-13]-Cu(20), where the layer thicknesses are noted in nm units in the parentheses and are known with an incertitude of $\pm 0.3\,$nm. The Cu layer was included to prevent oxidation of the Ho. The trilayers were deposited at ambient temperature by d.c. magnetron sputtering in $1.5\times 10^{-2}$ mbar of Ar onto oxidized Si substrates. The base pressure of the chamber was $10^{-9}\,$mbar, with an O$_2$ partial pressure $<10^{-11}\,$mbar. The samples were mounted on a rotating sample holder, whose angular velocity below the magnetron determines the thickness of the deposited layer: all the samples in each series were therefore deposited without breaking the vacuum \cite{JasonPRB}. While the growth conditions of all the samples in one run are exactly the same, the only difference being the monotonically increasing Ho thickness, variations between different runs can be expected due to changes in the residual oxygen pressure.
\begin{figure}[t]
\begin{center}
\includegraphics[width=7.5cm]{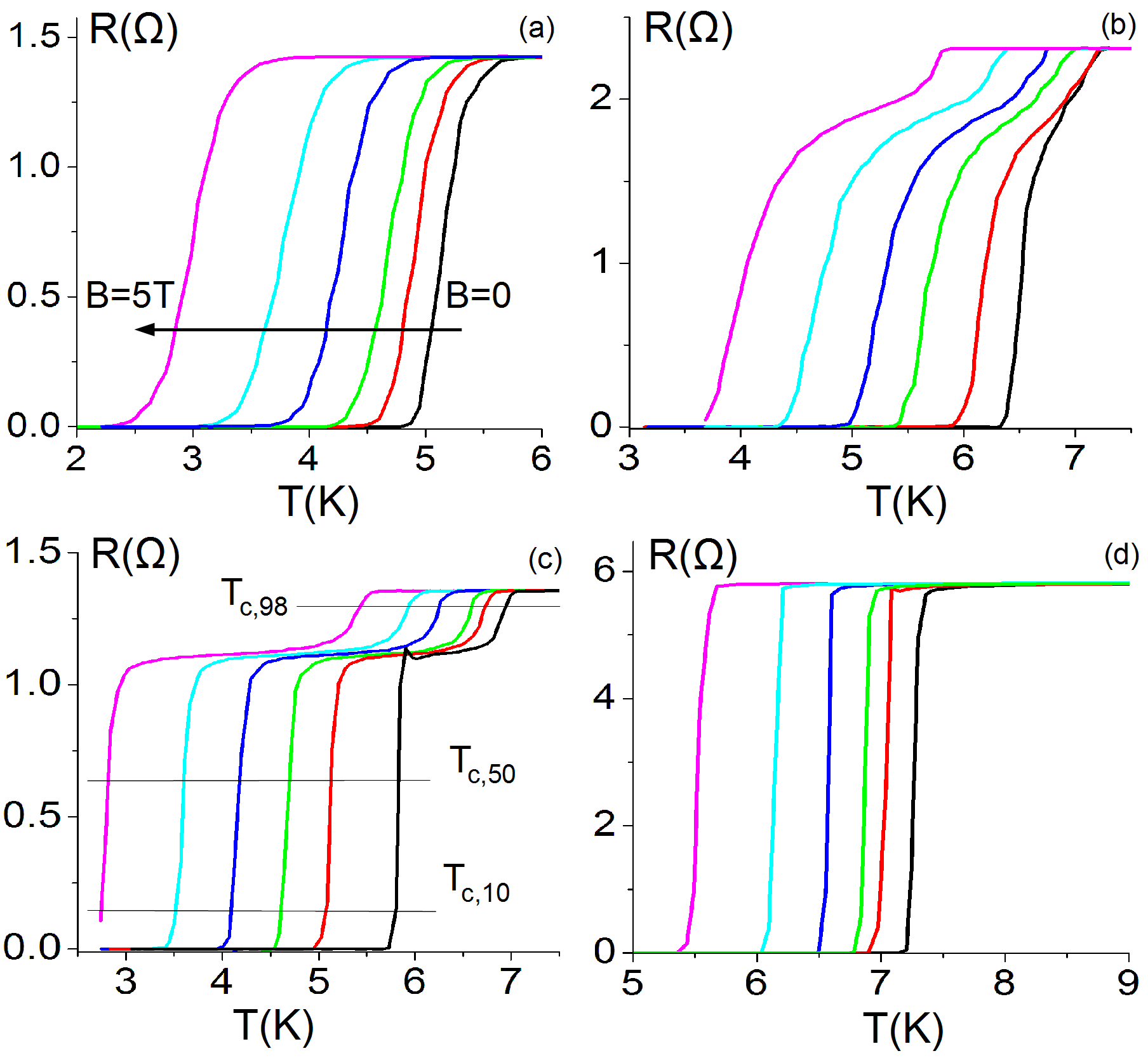}
\caption{(Color online) $R(T)$ curves with different in-plane fields applied $B=0,1,2,3,4,5 \,$T (from right to left, see arrow) for $d_{Ho}=4.4\,$nm (a), $d_{Ho}=5.4\,$nm (b), $d_{Ho}=7.4\,$nm (c) of T1, and for a control Nb sample (d).} 
\label{RT}
\end{center}
\end{figure}
The film resistance ($R$) as a function of temperature ($T$) was measured in a four-point configuration with an alternating current $I=\pm 50 \mu$A. The $R(T)$ curves at zero magnetic field were measured in a liquid $^4{\rm He}$ dip probe down to $T=4.2\,$K, while the $R(T)$ curves in the field range $0-5\,$T were measured in a pumped $^4 {\rm He}$ cryostat whose minimum temperature was $1.7\,$K. \\

\begin{figure}[t]
\begin{center}
  \includegraphics[width=8.5cm,height=4.7cm]{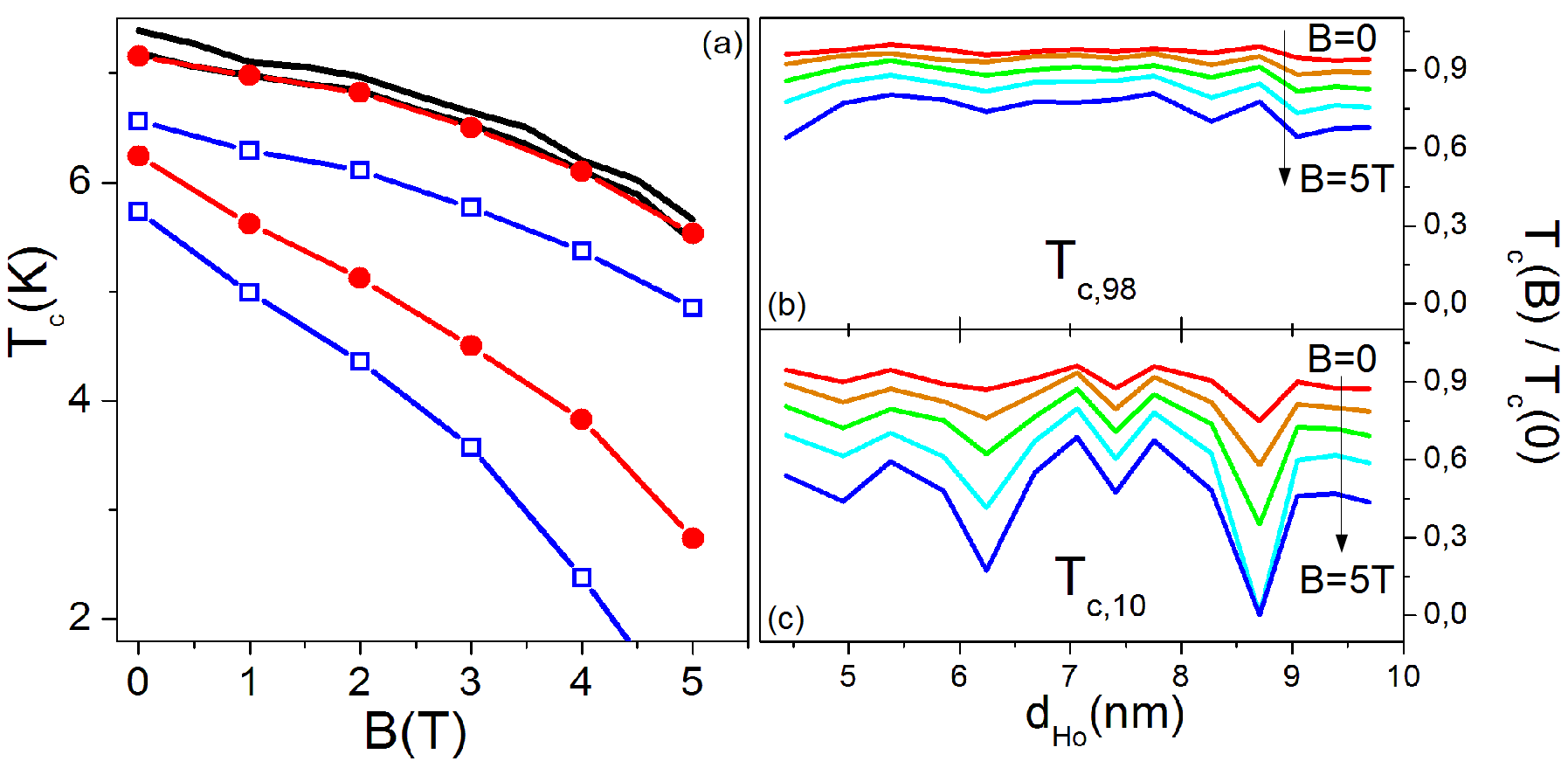}
  \caption{(Color online) (a) $T_{c,10}(H)$ and $T_{c,98}(H)$ for a Nb thin film (black lines, higher critical temperatures), and for two representative samples of the T1 series: $d_{Ho}=5\,$nm (red dots) and $d_{Ho}=6.2\,$nm (blue squares). Relative decrease with field of $T_{c,98}$ (b) and $T_{c,10}$ (a) as a function of Ho thickness for B=0,1,2,3,4,5 T from top to bottom. The scale is exactly the same in (b) and (c).}
\label{TcH}
\end{center}
\end{figure}

\emph{Results --} The $T_c$ values are extracted from the $R(T)$ curves; representative data are shown in figs.~\ref{RT}(a)-\ref{RT}(c) for three samples with in-plane magnetic field applied in the 0-5 T range. Two different cases are observed when varying the Ho layer thickness: either the transition is sharp with a width below $0.4\,$K [fig.~\ref{RT}(a)], or two transitions are present with a separation as large as $1.3\,$K at $B=0$ [fig.~\ref{RT}(c)]. To exclude the possibility that the double transition is due to inhomogeneity in the Nb layer thickness, we have deposited a 15-nm-thick control sample of Nb. Its superconducting transition [fig.~\ref{RT}(d)] is sharp with a transition width of $\lesssim 0.15\,$K. Control Ho films, deposited over thin (4nm) Nb layers also show a monotonic $R(T)$ curve in the experimental temperature range \cite{supp}. 
The complex transitions observed are therefore due to the interaction of the superconducting layer with the magnetic Ho layer: the films are structurally homogeneous, the heterogeneous electrical properties being seemingly generated by the heterogeneous magnetic structure. To follow the dependence of the two transitions observed, we record the temperatures corresponding to $98\%$, $50\%$ and $10\%$ of the normal state resistance ($R_N$) measured at $8.5\,$K (labeled $T_{c,98}$, $T_{c,50}$ and $T_{c,10}$ respectively). $T_{c,10}$ and $T_{c,50}$ have similar values, and correspond to the stronger proximity suppression of $T_c$, leading to the zero-resistance superconducting state. $T_{c,50}$ is the critical temperature of the transition responsible for more than 80$\%$ of the resistance drop and in many samples the only transition present in the R(T) curves. Moreover, as we will show in fig.~\ref{TcH}, $T_{c,50}$ is strongly affected by the magnetic field  'unzipping' the Ho spiral magnetisation. In fig.~\ref{Tc} we show $T_{c,50}$ as a function of $d_{Ho}$ for T1 and T2. In both series $T_{c,50}$ oscillates, with a period $\Delta d_{Ho} =1.8\pm 0.5\,$nm for T1 and $\Delta d_{Ho} =1.5 \pm 0.5\,$nm for T2. The amplitude of the oscillations can be as large as $0.5\,$K in T1 and $1\,$K in T2, and is much larger than the estimated error in the $T_c$ measurement (indicated on the left-most data point in each case). The oscillations are superimposed on an exponential decay $\exp(- d_{Ho}/l)$ with characteristic length scales $l = 47\,$nm for T1 and $l= 33\,$nm for T2. \\

\emph{Oscillatory $T_c$ --} In our Nb-Ho bilayers we can observe multiple oscillations, superimposed on a slow decay, which could correspond to the beginning of a reentrance similar to those observed in homogeneous F-S films 
(see also the larger range data presented in \cite{supp}). The clearer oscillations in T2 are possibly induced by the thinner Nb layer, on which the Ho has a stronger effect, leading to smaller $T_c$ and larger oscillation amplitudes \cite{Zdravko}.
\begin{figure}[t]
\begin{center}
  \includegraphics[width=\columnwidth]{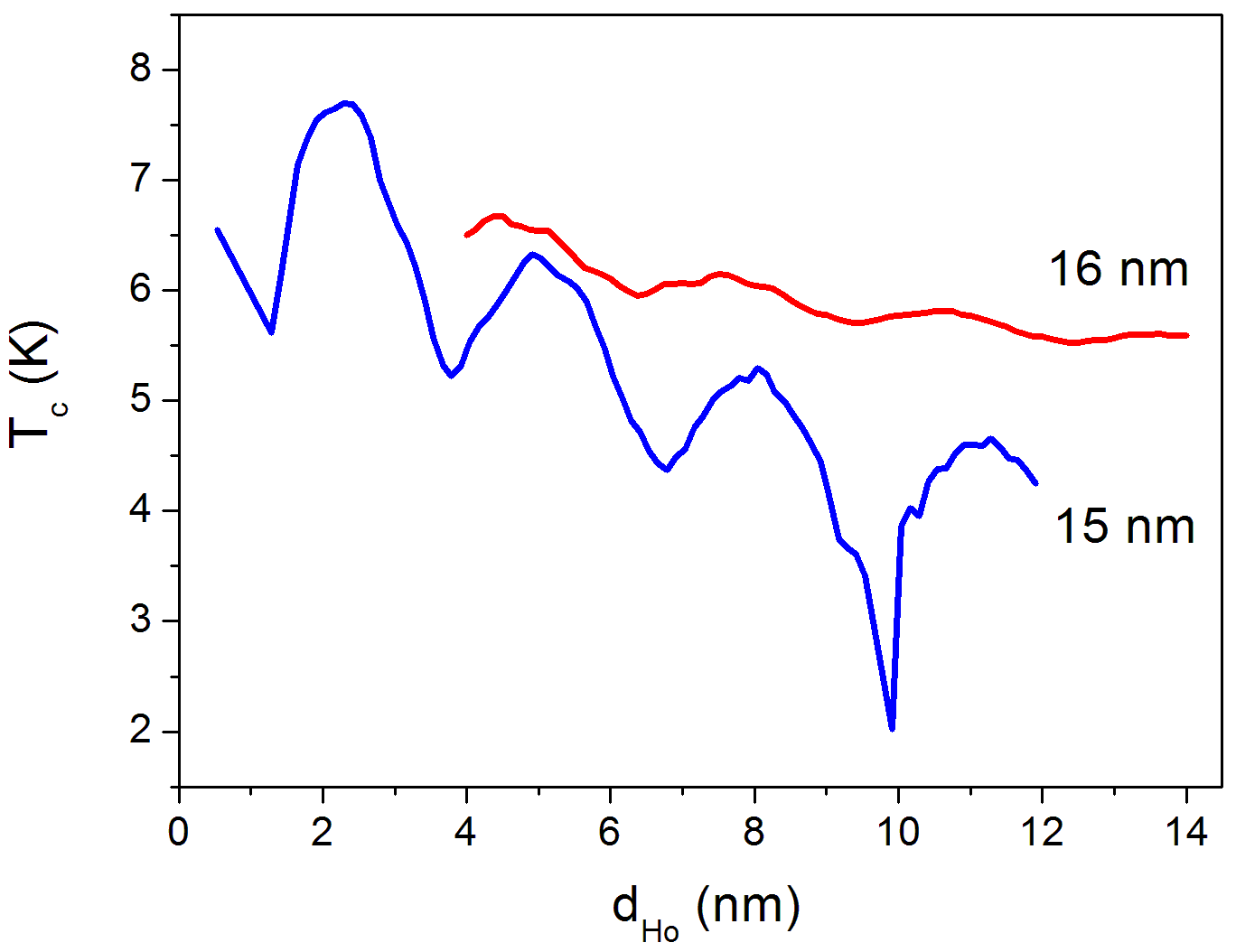}
  \caption{(Color online) Solution of the BdG equations for the T1 series (red, $d_s=16\,$nm) and the T2 series (blue, $d_s=15\,$nm). The theoretical curve for T1 was calculated using $\xi_{Ho}=3.2 \,$nm and an overall $T_c$ shift of $1.4\,$K; the curve for T2 was calculated using $\xi_{Ho}=2.1 \,$nm and an overall $T_c$ shift of $2\,$K. For both curves $T_c^0=9\,$K was used.}
\label{fit}
\end{center}
\end{figure}

To compare our results with theory, we have calculated the transition temperature of Nb-Ho bilayers using a fully microscopic procedure, based on the exact numerical diagonalization of the self-consistent Bogoliubov-de Gennes (BdG) equations \cite{pbov,ilya,ct}. This method can effectively handle the geometrical effects inherent to finite systems, as well as the triplet correlations generated by the presence of the nonuniform magnetization in Ho \cite{trip}. The $m=0$ and $m=\pm 1$ triplet amplitudes are all found to be nonvanishing, and similar in importance. 
The bilayers are theoretically modeled as being infinite in the transverse direction. The helical magnetic structure of Ho is described by a spiral local exchange field whose angle with the c-axis $\alpha$, turning angle $\beta$ and helical period $\lambda$ are given in the literature. 
In general, whenever possible, we set the relevant parameters to their corresponding experimental values, keeping the adjustable ones to a minimum. Thus, we set the Ho and Nb layers thicknesses $d_{Ho}$ and $d_s$, and the critical temperature $T_c^0$ of a bulk Nb sample, to their measured values. In the thin Nb layer, we set the effective correlation length and the average Fermi velocity to standard values, respectively $\xi_s=10\,$nm and $v_s= 4.4 \times 10^5\,$m/s \cite{Nb}. The dimensionless Ho exchange field $h_0$, which determines the ferromagnetic coherence length $\xi_{Ho}$, is an adjusted parameter.
The calculated $T_c$ is shown in fig.~\ref{fit} as a function of $d_{Ho}$. To account for the measured magnetic dead layer, an horizontal shift $\sim 1\,$nm has to be applied when comparing to the experimental data \cite{James,supp}. We obtain, as in the experiments, multiple oscillations, with a similar $T_c$ decay and oscillations amplitude, broadly reproducing the experimental shape of $T_c$ as a function of $d_{Ho}$. 
Although the absolute value of $T_c$ differs (by $\sim 1.5\,$K, see caption of fig.~\ref{fit}), such shifts have been previously seen in \cite{ilya} due to the actual samples not quite conforming to the ideal assumptions of the theory. We find that the results are very sensitive to the value of $d_s$, making it therefore remarkable that reasonable agreement is found without adjusting this parameter from its experimental value. The $d_{Ho}$ range of the T2 series is rather extensive and includes the small $d_{Ho}$ range where the $T_c$ oscillations are more pronounced. Of particular interest is the sharper minimum found near $10\,$nm. According to the theory this is due to the near emergence of a reentrance region in $d_{Ho}$: the system is very close to being non-superconducting at this thickness value, as found experimentally (see arrow in fig.~\ref{fit}). 
The adjusted values of the parameter $h_0$ used here give $\xi_{Ho}=3.2 \,$nm for the T1 series and $\xi_{Ho}=2.1 \,$nm for the T2 series. These are both consistent with $\xi_{Ho}=4.3\,$nm found experimentally \cite{James}. Due to the ratio $ \xi_{Ho}/\lambda \sim 1 $, the oscillations in $T_c$ are governed by a combination of both the helical period $\lambda$ and the ferromagnetic coherence length $\xi_{Ho}$, leading to an intricate global behavior. 
%Overall, this intricate behavior is remarkably similar to that of the experimental results. The differences may be explained by the polycrystalline nature of the films, not included in the theory. Indeed, the presence of the (101) phase may influence the magnetic repeat distance and thus the $T_c$ oscillations periodicity, which is determined by both the ferromagnetic coherence length and the Ho helical period. 
While to provide a definitive test of the theory requires a knowledge of the exact magnetic structure, our primary aim is to demonstrate that our unusual results are at least consistent with a theory based on bulk Ho properties. \\

\emph{$R(T)$ dependence --} We now consider the two separate transitions appearing in the $R(T)$ curves. To try to understand their origin, we have (i) applied a magnetic field up to $5\,$T and (ii) reduced the lateral dimensions of the films down to $160\,$nm. As a function of the Ho thickness, we observe different behaviors as in fig.~\ref{RT}(a)-(c). 
If the transition is a single step at $B=0$, the magnetic field can either act only as a pair-breaking mechanism, resulting in a progressive decrease of $T_c$ [fig.~\ref{RT}(a)], or it can split the transition in two resistance drops, whose distance in temperature increases with the field amplitude [fig.~\ref{RT}(b)]. If two transitions are already present at $B=0$, they will grow apart with the magnetic field [fig.~\ref{RT}(c)]. In all cases, the amplitude of the two resistance drops is not affected by the magnetic field. When applying an out-of-plane magnetic field, the result is qualitatively the same, but the $T_c$ suppression is more than twice as strong. We have compared the in-plane magnetic field dependence of the two transitions with the field dependence of a control Nb layer. In fig.~\ref{TcH} (a) we trace $T_{c,10}$ and  $T_{c,98}$ for the control Nb layer and two representative bilayers from the T1 series.
While $T_{c,98}$ follows the dependence of the control Nb dependence with field, even though with a lower $T_c$,
%albeit with a sample dependent maximum $T_c$,
 $T_{c,10}$ is suppressed on average 1.5 times faster. 
Fig.~\ref{TcH} (b) and fig.~\ref{TcH} (c) show the relative decrease of $T_c$ with the magnetic field for the two transitions ($T_{c,98}$ and $T_{c,10}$ respectively) as a function of the Ho thickness. As already observed in fig.~\ref{RT}, the magnetic field has different effects over the two transitions: on average $T_{c,98}$ is less affected by the field and its decrease is quite homogeneous over all the samples, while $T_{c,10}$ decrease is larger and, most importantly, strongly dependent on the Ho thickness.

%The width of the step $\Delta T_c$ is shown as a function of Ho thickness and magnetic field in Fig.~\ref{TcH}(b). As already observed in Fig.~\ref{RT}, the magnetic field increases the width of the step, which also depends strongly and non-monotonically on the Ho thickness.
We can thus identify two regions in our bilayers: in one, proximity superconductivity can be easily induced in the Ho layer, and the magnetic field affects similarly all the samples through the reduced $T_c$ of Nb; the other becomes superconducting at a substantially lower temperature, and the effect of the magnetic field is not limited to the Nb layer but also affects the magnetic structure for certain Ho thicknesses. The difference between the two transitions can be as high as $1.3\,$K, and is increased by the magnetic field, attaining in some samples $4\,$K. Since such two-phase behavior is not present in the Nb layer, we attribute its origin to the Ho layer and its magnetic structure. It can be explained if we assume the presence of magnetically-inactive regions which give rise to a higher temperature transition. Since the roughness is much smaller than the layer thicknesses \cite{supp}, the Ho films are homogeneous, while the magnetic structure is not - either because of the presence of misoriented grains or because of domain structures. The size of the domains has been measured in epitaxially grown Ho to be $\sim 50\,\mu $m \cite{Hodom}, but we expect it to be smaller in our polycrystalline films. To explore the spatial dependence of the phase separation, we have drawn tracks on the film and made constrictions of fixed length $L=10\,\mu$m and variable width $w$ in the $160\,$nm - $2\,\mu$m range, by Focused Ion Beam etching. In the constrictions, the $R(T)$ curves are qualitatively similar to those measured in the films, presenting two separate transitions at exactly the same critical temperatures as in the films, but with a varying, lower, step resistance (see \cite{supp}). This suggests that we can modify the proportion of the two phases without affecting their electronic properties when the device size approaches the characteristic size of the regions in which the phases are confined.\\

%This suggests that the current percolation path, and possibly the morphology of the regions associated with the different transitions, are changed by reducing the dimensionality. \\

\emph{Conclusions --} In qualitative agreement with the theoretical modeling of inhomogeneous F-S junctions, our experiments show multiple complex oscillations of $T_c$ vs $d_{Ho}$ resulting from both the contributions of singlet and triplet Cooper pairs, and controlled by the helical magnetism and the ferromagnetic coherence length in Ho. Even though the model is based on a monocrystalline structure, the agreement is surprisingly good, giving evidence of interesting and potentially important behaviour which will hopefully stimulate further experimental and theoretical research. The presence of a behavior which is already quasi reentrant in $d_{Ho}$, and which is in agreement with the theory, is particularly intriguing and will be addressed in future work. We furthermore observe that the $R(T)$ curves show for certain Ho thicknesses two transitions, affected differently by the magnetic field. Further work is thus in progress to define the nature of the two phases in which the proximity effect acts so differently.\\

%The data suggest that our samples are magnetically inhomogeneous, and are compatible with a picture in which magnetic domains and domain walls play a key role. However, the picture is not simple, given that even large applied fields (which would be expected to significantly modify the magnetic microstructure) do not alter the relative proportions of the two $T_c$ phases. Further work is thus in progress to define the nature of the two phases, and how they are affected by the proximity effect. 

This work was funded by the UK EPSRC and the Royal Society through a University research Fellowship. O. T. Valls and C. T. Wu thank IARPA grant N66001-12-1-2023 for partial support.

\end{document}

\title{Supplementary Materials for: 'Supra-oscillatory critical temperature dependence of Nb-Ho bilayers'}
%\author{A.D. Chepelianskii$^{(a,b)}$ , D Klinov$^{(c)}$ , A Kasumov$^{(b)}$ , S Gu\'eron$^{(b)}$ ,
%O Pietrement$^{(d)}$ , S Lyonnais$^{(e)}$ and H Bouchiat$^{(b)}$ \\
%(a) Cavendish Laboratory, University of Cambridge, J J Thomson Avenue, Cambridge CB3 OHE, UK \\
%(b) LPS, Univ. Paris-Sud, CNRS, UMR 8502, F-91405, Orsay, France\\
%(c) Shemyakin-Ovchinnikov Institute of Bioorganic Chemistry, Russian Academy 
%of Sciences, Miklukho-Maklaya 16/10, Moscow 117871, Russia \\
%(d) UMR 8126 CNRS-IGR-UPS, Institut Gustave-Roussy, 39 rue Camille
%Desmoulins, 94805 Villejuif Cedex, France \\
%(e) Museum National dâHistoire Naturelle, CNRS, UMR7196, Inserm, U565,
%43 rue Cuvier, 75005 Paris, France
%}

\author{F. Chiodi$^1$, J. D. S. Witt$^2$, R. G. J. Smits$^1$, L. Qu$^1$, G. B. Hal\'asz$^1$, C.-T. Wu$^3$,
O. T. Valls$^3$, K. Halterman$^4$, J. W. A. Robinson$^1$, and M. G. Blamire$^1$ \\
$^1$ \small{\itshape{Department of Materials Science, University of Cambridge, Pembroke Street, Cambridge CB2 3QZ, UK}} \\
$^2$  \small{\itshape{School of Physics and Astronomy, E. C. Stoner Laboratory, University of Leeds, Leeds LS2 9JT, UK}}\\
$^3$ \small{\itshape{School of Physics and Astronomy, University of Minnesota, Minneapolis, Minnesota 55455}}\\
$^4$ \small{\itshape{Michelson Lab, Physics Division, Naval Air Warfare Center, China Lake, California 93555}} \\
}

%\author{F. Chiodi}
%\email{francesca.chiodi@u-psud.fr}
%\altaffiliation{also at Institut d'Electronique Fondamentale, CNRS-Université Paris-Sud, 91405 Orsay, France}
%%\affiliation{{\mbox{Department of Materials Science and Metallurgy,University of Cambridge, Pembroke Street, Cambridge CB2 3QZ, United Kingdom}}}
%\author{J. D. S. Witt}
%%\affiliation{{School of Physics and Astronomy, E. C. Stoner Laboratory, University of Leeds, Leeds LS2 9JT, United Kingdom}}
%\author{R. G. J. Smits}
%\author{L. Qu}
%\author {G. B. Hal\'{a}sz}
%\affiliation{{\mbox{Department of Materials Science and Metallurgy,
%University of Cambridge, Pembroke Street, Cambridge CB2 3QZ, UK}}}
%
%\author{\mbox{C.-T. Wu}} 
%\author {O. T. Valls} 
%\altaffiliation{Also at Minnesota Supercomputer Institute, University of Minnesota,
%Minneapolis, Minnesota 55455}
%\affiliation{{School of Physics and Astronomy, University of Minnesota, Minneapolis, Minnesota 55455}}
%
%\author{K. Halterman}
%\affiliation{{Michelson Lab, Physics Division, Naval Air Warfare Center, China Lake, California 93555}}
%
%\author{J. W. A. Robinson} 
%\email{jjr33@cam.ac.uk}
%\author{M. G. Blamire} 
%\affiliation{{\mbox{Department of Materials Science and Metallurgy,
%University of Cambridge, Pembroke Street, Cambridge CB2 3QZ, United
%Kingdom}}}


\begin{abstract} 
In this Supplementary Materials we provide the magnetic and structural characterization of our sputtered polycrystalline Nb/Ho/Cu trilayers. Moreover, we show supplementary $T_c$ vs. Ho thickness data in which we explore the high Ho thickness range. Finally, we show the effect of a lateral constriction on the films' $R(T)$ curves.
\end{abstract}

\maketitle

\section{Ho magnetic properties}

\begin{figure}[hb]
  \includegraphics[width=0.9\columnwidth]{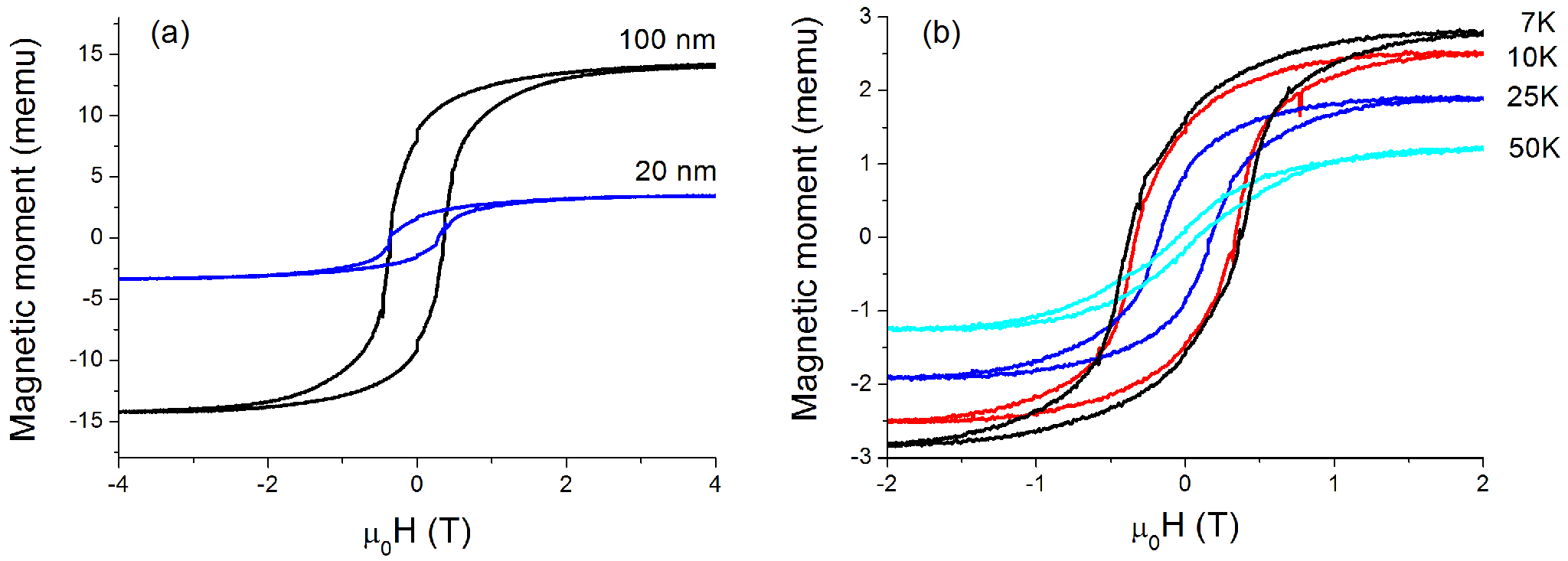}
  \caption{(Color online) (a) Magnetic moment vs magnetic field at T=10 K for the two samples Nb(4)/Ho(20)/Au(10) and Nb(4)/Ho(100)/Au(10), where the Au layer protects the Ho from oxidation (the layer thicknesses are noted in nm units in the parentheses). The small jump at zero field is a measurement artifact. (b) Magnetic moment vs magnetic field for the Nb(4)/Ho(20)/Au(10) sample at T=7,10,25,50 K. The slightly higher value of saturation moment in (a) is due to a better subtraction of the diamagnetic background.}
\label{MH}
\end{figure}

The magnetization of the Nb-Ho films was determined from vibrating sample magnetometry at varying temperature for two Ho layer thicknesses, $d_{Ho}=20\,$nm and $d_{Ho}=100\,$nm [Fig. \ref{MH} (a)]. In all measurements the applied field was in-plane and along the long axis of the sample. %The $M(H)$ loops all have low coercivity and low remnant magnetization indicating that the Ho is antiferromagnetically ordered. 
The films become saturated at $B\approx 4\,T$. Extrapolation from the plot of the saturation magnetization as a function of Ho layer thickness reveals that the Ho has a magnetically dead-layer of $\approx 1\pm 0.5\,nm$ per Ho surface. The determined magnetisation is $2.7\times 10^6$ A/m, lower than the theoretical value of $3.08\times 10^6$ A/m, suggesting that magnetic inhomogeneity is present in our sputtered polycrystalline films. These measurements are in good agreement with the characterization of ref. \cite{James}.\\

\newpage

\section{Samples characterization}

\begin{figure}[ht]
  \includegraphics[width=0.7\columnwidth]{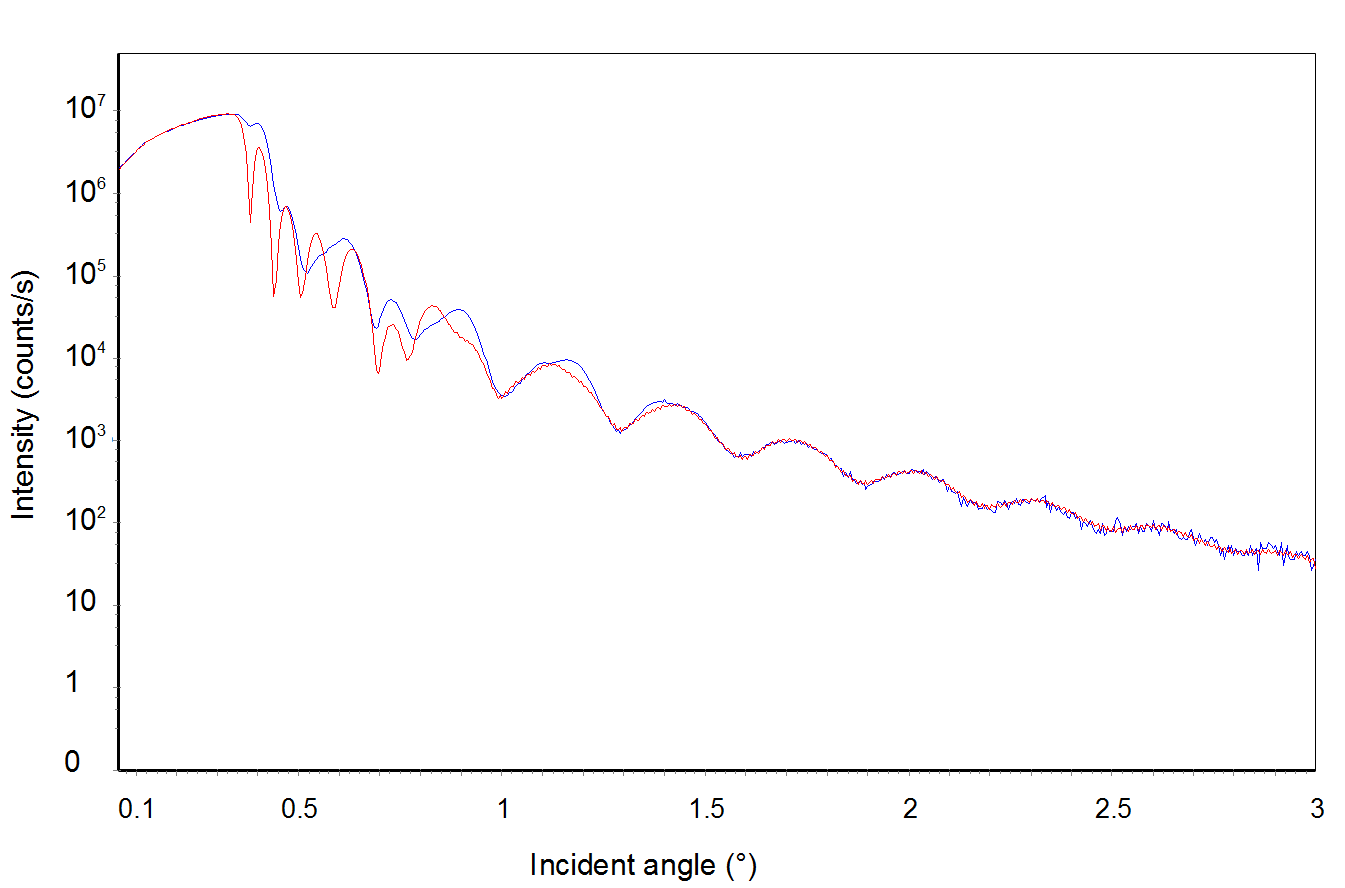}
  \caption{(Color online) Low angle X-ray reflectivity for $d_{Ho}=5\,$nm of the T1 series (blue, higher curve). The theoretical curve (red, lower curve) is calculated for the multilayer Nb(14.4)/Ho(5)/Cu(17)/CuO(2.3) on a Si/SiO$_2$ substrate (the layer thicknesses are noted in nm units in the parentheses). The sample roughness of 1.7$\,$nm is mainly due to the CuO layer, while the roughness at the Nb level is 1$\,$nm. The disagreement at the lower angles may be due to little Nb oxidation at the Nb/SiO$_2$ interface. This, and the formation of a 1 nm dead layer, account for the lower Nb thickness found in the X-rays measurements.}
\label{xrays}
\end{figure}

Deposition rates and film thicknesses have been checked on pre-deposited step-edges and from low-angle X-ray reflectivity on T1 and T2 samples (Fig. \ref{xrays}); the latter gives an estimation of the film roughness of about $1.5\,$nm. \\

Analysis of the integrated intensities of high-angle X-ray diffraction data taken from a thick bilayer film reveals that the Ho exhibits texturing with (002) as the primary phase, the c-axis being normal to the sample surface; a secondary (101) phase and a much weaker (100) tertiary phase are also present \cite{James}. \\

The Residual Resistance Ratio (RRR), the ratio of the resistance at 292K and 10K, was also measured to estimate the films quality. We obtained RRR$\sim$1.3, with a 20$\%$ variation from sample to sample. The samples of the two series T1 and T2 have similar RRR values, as expected from using the same fabrication method.\\

\newpage

\section{Resistance vs Temperature for Ho samples}

Control Ho films, deposited over thin (4nm) Nb layers, show a monotonic $R(T)$ curve in the experimental temperature range. Since neither the Ho films nor the Nb films present any heterogeneous behaviour, the double transition observed in the Nb/Ho bilayers can be attributed to the effect of the Ho magnetism on the Nb proximity effect. 

\begin{figure}[ht]
  \includegraphics[width=0.5\columnwidth]{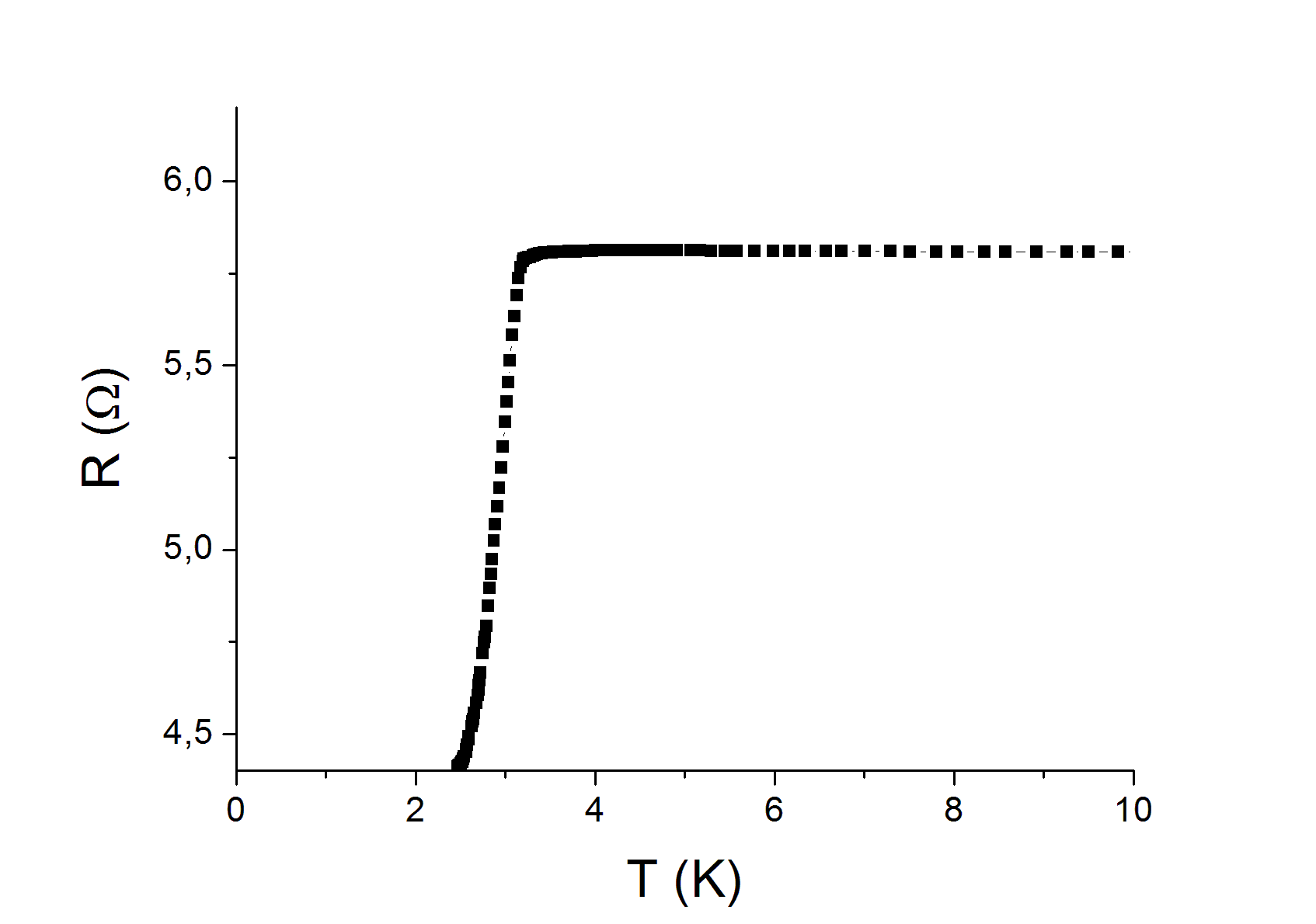}
  \caption{(Color online) Resistance vs. temperature for the Nb(4)/Ho(20)/Au(10) sample. The resistance drop at T=3 K is due to the superconducting transition of the thin Nb layer. Note that in the temperature measurement range 4-8 K, the resistance of the Ho layer is extremely stable.  }
\label{RT}
\end{figure}

\section{Larger Ho-thickness range data}

To explore a larger range of Ho thicknesses, we have deposited Nb(15)/Ho(x)/Cu(20) trilayers with Ho thickness in the 4-32 nm range. The oscillations are here superimposed on a slow oscillation with a minimum around $d_{Ho}=21\,$nm. This confirms the hypothesis that the slow decay seen in series T1 and T2 is but the initial part of a slow oscillation similar to that seen in homogeneous S-F bilayers. 

\begin{figure}[hb]
  \includegraphics[width=0.5\columnwidth]{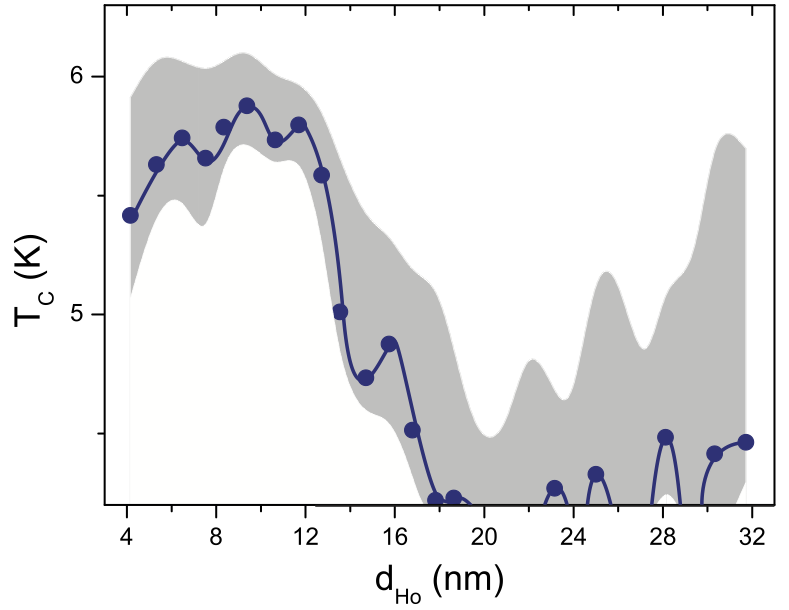}
  \caption{(Color online) Superconducting critical temperature $T_{c,50}$ vs. Holmium thickness $d_{Ho}$ of a larger range series, fabricated in the same way as T1 and T2. The curves which delimit the shaded area are $T_{c,10}$ and $T_{c,98}$. }
\label{Tcho1}
\end{figure}

\newpage

\section{Resistance vs Temperature for constrictions of different width}

We show here the $R(T)$ renormalized curves for a film in which constrictions have been created (see inset). While the amplitude of the two transitions is strongly and non-monotonously affected by the decrease of the lateral dimensions, we observe that the transitions temperatures correspond to those of the initial film.

\begin{figure}[ht]
  \includegraphics[width=0.5\columnwidth]{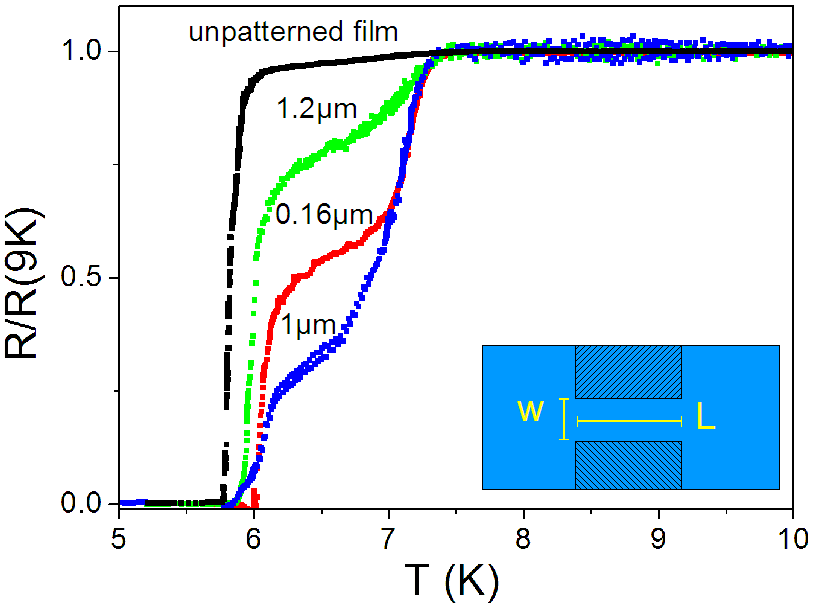}
  \caption{(Color online) Renormalized resistance vs. temperature for $d_{Ho}=7.75\,nm$ of T1. From top to bottom: film and constrictions with $L=2\,\mu m$ and $w=1200\,nm$, $w=160\,nm$ and $w=1000\,nm$.   }
\label{RTw}
\end{figure}